\documentstyle[psfig,epsf]{texas}
\def\lsim{\lower.5ex\hbox{$\; \buildrel < \over \sim \;$}}
\def\gsim{\lower.5ex\hbox{$\; \buildrel > \over \sim \;$}}
\begin{document}
\title{Modelling the Origin of Astrophysical Jets from Galactic and
Extra-galactic Sources}
\noindent
{\it A New Approach to Combine the Accretion-Wind Topologies}
\author{Tapas Kumar Das}
\address{S. N. Bose National Centre For Basic Sciences\\
Block JD Sector III Salt Lake Calcutta 700 091 WB India\\
\rm Email: tdas@boson.bose.res.in}
\begin{abstract}
Though widely observed to be emanating out from a variety of galactic and
extra-galactic sources, the underlying physical mechanism behind the
formation of the cosmic jets and outflows are wrapped
around by the veil of mystery till date. Neither the amount of matter
contained in these jets could accurately be calculated by any
definitive method. In this paper we present a theoretical model, which,
{\it for the first time} we believe, is able to explain the jet formation
phenomenon as well as can compute the mass outflow rate by self-consistently
combining the exact transonic accretion-outflow topologies. Our model
could also analyze the dependence of this rate on various physical
parameters governing the inflow.
\end{abstract}
\noindent
{\large\bf Prologue}\\ \\
\noindent
Astrophysical jets are physical conduits along which mass, momentum,
energy and magnetic flux are channelled from the stellar, galactic
and extra-galactic objects to the outer medium. Geometrically these jets
are narrow (small opening angle) conical or cylindrical/semi-cylindrical
protrusions covering an astonishing range in size. While the jets
associated with young stars are typically $10^{17}$ cm in length,
jets from some giant extra-galactic sources have an overall extent in
excess of $10^{24}$ cm [1]. Thus the jet phenomenon is seen
on scales that cover more than seven orders of magnitude and some of
the extra-galactic radio jets are considered to be the largest 
single coherent structure found
in the universe. Though it is now an well-known fact that a variety
of celestial objets, spanning from stars (having all masses) during their
formation (Young Stellar Objects) to Active Galactic Nuclei 
\footnote
{`Active' galaxies are distinguished from ordinary galaxies in that they
show indications of having energy output not related to ordinary
stellar processes (standard thermonuclear evolution of energy). Their
`activity' is centered in a small nuclear region (R $<<$ 1 pc $\sim 3.1X10^{18}$
cm) and are associated with strong emission lines. Their power outputs
are dramatically enormous ($\sim 10^{46}$ ergs $s^{-1}$) and equal the
mass - energy equivalent to several solar masses per year [3]. The nucleus
of such galaxies are named ``Active Galactic Nuclei"(AGN).}
(and possibly sources of $\gamma$ ray bursts
also [2]) suffer mass loss through jets, the detailed nature
of the origin of the extragalactic jets is not quite clear due to the
lack of proper understanding of the underlying physical mechanism
responsible for jet production. In this article efforts have been made
to justify the validity of a model very recently proposed by us, which,
for the first time we believe, is able to explain the formation of
extragalactic cosmic jets using a self-consistent inflow-outflow
system. 
Primarily intended to make intelligible to readership not working
exclusively in this field, the philosophy behind the subject is 
essentially emphasized in this article instead of providing the 
mathematical details. This article, thus, is absolutely freed from
a single mathematical equation/formula. Readers interested 
to delve into the technical details are suggested to go through 
the references cited at appropriate places. \\
\section {\bf Introduction}
One of the most prominent signatures of activities around the Active
Galactic Nuclei (AGN)
is the presence of mass outflows and
jets. AGNs produce cosmic jets through which immense amount of matter
and energy are ejected out of the cores of the galaxies [3].
The structure of these jets can reach sizes of several
million light years and extent way beyond their host galaxies into the
vastness of intergalactic space. Similarly, micro quasars 
\footnote{Microquasars are stellar-mass black holes in our galaxy which 
mimic, in a smaller scale, of the phenomena seen in quasars. Unlike the
AGN and Qyasar jets (where the extend of the jets may reach several million 
of light years), the double sided jets coming out of these objets 
can have sizes upto a few light years only[4]}
have also been very recently discovered where the mass
outflows are formed from stellar mass black hole candidates [4].\\[0.25cm]
\begin{center}
\leavevmode
\epsfbox{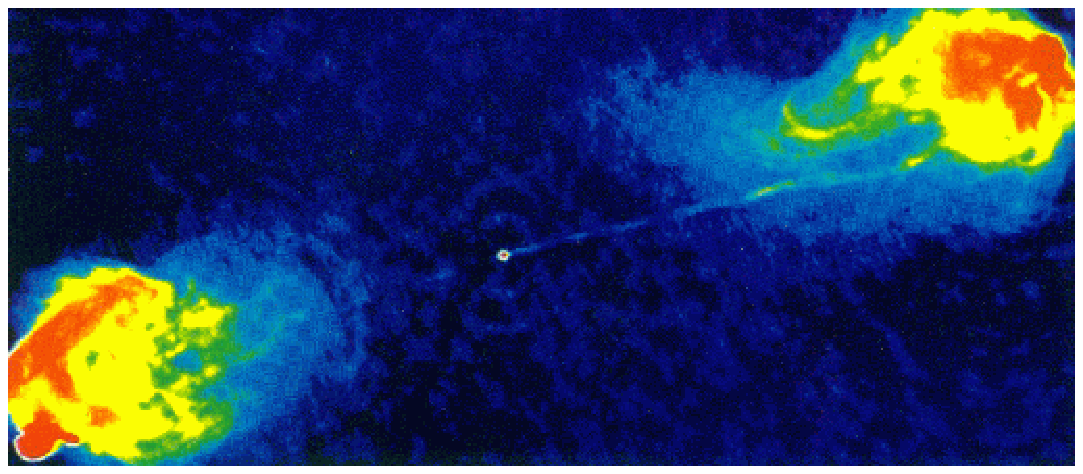}
\end{center}
\begin{center}
{{\bf Fig. 1:} {\small VLA radio image of Cygnas A after 
repeated correction from
atmospheric effects and calibration errors (Observation by R. H. Perley 
and J. W. Dreher). The bright spot in the centeris the core of the 
galaxy. Two bright patches on either side of the core (one at the top
right and the other at the lower left corner of the image) represent
the radio emitting lobes while the thin band connecting the central spot with
the top right patch is the jet shooting out from the core (Reproduced
with kind permission from R. H. Perley and J. W. Dreher)}}\\[0.25cm]
\end{center}
\noindent
Looking back at the era of infancy of radio observations of the
extragalactic sources, one sees that in 1953, Jenison and Dasgupta
[5] discovered that the radio emission from
Cygnas A was originating from two amorphous blobs straddling symmetrically
the associated optical galaxy rather than the galaxy itself
(Figure 1.). Subsequent
observations of other powerful radio emitting sources ravaeled the
fact that this is rather a general phenomenon. Initially it was thought
that these radio emitting blobs had been directly shot out of the core
of the galaxy, however, this idea created some dynamical problems
(adiabatic loss problem, as for example, see [3] for detail) which
prompted theoreticians to propose a ``black box model" to explain the
phenomenon. In early theoretical contributions, what people put
forward was basically a black box sitting at the dynamical
center of the galaxy which is doing something interesting so that these
radio emitting lobes are continuously fueled by the process of channelling
matter and energy emanating from the galactic center. There is need to go
into the history of subsequent discussion (and there is no scope to do
so either due to the limitation of space, interested readers may have a
look to [3] and [6-7]), suffice it to say
that from the present status of observational evidences [8],
we are in a strong position to say that these channels
of matter and energy, or jets (as it was first named by Baade and
Minkowski in 1954 [9]) are the ubiquitious
feature of the AGNs, Young Stellar Objects (YSOs) and of some small scale
prototype of AGNs, SS433 for example, which is believed to harbour
a neutron star at its center. (For a detail discussion of
SS433 jet, see [10-11]).
Although since the first theoretical contribution to this black box
model approach [12], much work has been done
on how such jets interact with their surrounding and on how such interaction
may convey the informations about the morphology of different
extragalactic radio sources, the fundamental problem of what 
{\it exactly} is happening inside the black box still remains unresolved.\\
\noindent
>From the observational point of view, probably
the most attractive feature of these astrophysical jets is that they
are the most prominent and visible signatures of the AGNs. Hence, studying
the jets has been one of the most exhaustive part of the research carried
out by the observational astrophysisists over many years, and a huge ``zoo" of different jet
species has emerged.\\
\noindent
On the other hand, from the theoretical front, the non-stellar activities
around the AGNs are thought to be produced by a powerful engine
sitting at the dynamical center of the galaxy [13].
Because of the fact that the high luminosity produced by the AGNs are
concentrated in a very small volume, it has been strongly argued that
these engines are basically powered by the accretion onto massive
black holes. This ``black hole hypothesis", namely
that essentially all AGNs contain $\sim 10^6 - 10^9 {M_{\odot}}$ black holes
[13-14] and that these objects together with
their orbiting accretion disks are the prime movers for most  of the powerful
activities of AGNs including the formation of bipolar outflows and relativistic
jets [15] are further supported by the recent observational
evidences, specially from the VLBI observations and the HST data where the
signature of the so called ``jet disk symbiosis" is supposed to be detected
[16].
That means, for most (if not all) AGNs and microquasars, the jets and the
accretion disks around the central compact object are symbioticaly
related [17]. Probably this has to be the case in reality
because in the absence of any binary companion, jet is supposed to be
the only outlet for the intrinsic angular momentum of the 
interstellar/ intergalactic matter accreting onto an isolated compact
object. So the accretion powered outflows
are not merely an incidental by-product of the mass flow through the
disk but, in fact, are a necessary ingredient in the accretion process,
in that they constitute the main mechanism for removing excess angular
momentum of the inflowing matter. Hence, it is quite logical to conclude
that the jet formation and accretion onto isolated black holes are not
two different issues to be studied disjointedly, but they must be
strongly correlated and it is {\it necessary} to study the accretion and jet
within the same framework. On the other hand, the major difference
between the ordinary stellar outflows and the outflows/jets from the vicinity
of a black hole or a neutron star, is that they do not have their
own atmospheres and outflows/jets in this case {\it have to be generated from
the inflowing materials only}.\\
\noindent
Keeping these basic facts in the back of our mind, our aim
was to theoretically study the mass outflow from galactic/
extragalactic sources more realistically than what has been attempted
so far. The existing models which study the origin, acceleration and 
collimation of mass outflow in the form of jets from AGNs and Quasars 
are roughly of three types. The first type of solutions confine 
themselves to the jet properties only, completely decoupled from the 
internal properties of accretion disks [18-20].
In the second type, efforts are made to 
correlate the internal disk structure with that of the
outflow using hydrodynamic, 
magnetohydrodynamic or electromagnetic  considerations ([17] and
references therein, [21-24]). 
In the third type, numerical simulations are carried out
to actually see how matter is deflected from the equatorial plane
towards the axis [25-31]. 
>From the analytical front, though the wind type and accretion
type solutions come out from the same set of governing
equations [23-24], till today there was no attempt
to obtain the estimation of outflow rate from the inflow rate. 
A theoretical model has very recently been developed [32-40],
which, for the first time 
we believe, can compute
(semi analytically and semi numerically) the absolute value of mass
outflow rate from the matter accreting onto galactic and extra-galactic
black holes using combinations of exact transonic accretion and wind
topologies which form a self-consistent inflow-outflow system. 
The simplicity of black holes and neutron stars lie in the fact that they do
not have atmospheres. But the inflowing matter surrounding them have, 
and similar method
as employed in stellar atmospheres should be applicable to the
accreting matter surrounding them. 
The approach in our model is precisely this. We first
determine the properties of the inflow and outflow
and identify solutions to connect them. In this manner we
self-consistently determine what fraction of the matter accreting onto
these compact objects is coming out as outflows/jets. \\
If ${{\dot M}_{in}}$ is the time rate of accretion onto a compact 
object (the amount of matter {\it falling in} per unit time) 
and ${{\dot M}_{out}}$
be the time rate of outflow (the amount of matter being `kicked out'
per unit time as {\it wind}), the 
ratio ${\left(\frac{{\dot M}_{out}}{{\dot M}_{in}}\right)}$ we call the 
`Mass Outflow Rate' and denote it by $R_{\dot m}$. Thus $R_{\dot m}$
is a measure of the ratio of the outflow rate to the inflow rate of
matter. The major aim of our work is to compute the absolute value 
of $R_{\dot m}$ in terms of the inflow parameters and to study the
dependence of $R_{\dot m}$ on those parameters. Computation
has been carried out for a Schwarzschild black hole using
Paczy$n^{\prime}$ski-Wiita [41] pseudo-Newtonian potential
which mimics the Schwarzschild space-time in an excellent fashion.\\
For matter accreting with considerable intrinsic angular momentum (formation
of accretion disks), we establish [32-36] that the bulk 
of the
outflow is from the CENtrifugal pressure dominated BOundary Layer
(it is called CENBOL, the formation of which will be explained in the 
next section), we find that $R_{\dot m}$ 
varies anywhere from a few percent to even close to a hundred
percent depending on the initial parameters of the inflow, the degree
of compression of matter near the CENBOL and the polytropic index of the
flow. Our model thus, not only provides a sufficiently plausible
estimation of $R_{\dot m}$, but is also able to study the variation of
this rate as a function of various parameters governing the flow [33-36].\\
We have also studied the mass outflow from spherical
accretion with zero intrinsic angular momentum [37-40]. It has been
shown that a self-supported spherical pair-plasma mediated standing shock
may be produced even for accretion with zero angular momentum. We have taken
this shock surface as the generating surface of mass outflow for
spherical inflow and have compared the results with that of obtained from the
outflows generating from CENBOL (disk-outflow system).\\
The plan of this article is as follows:\\
\noindent
In the next section we describe our model for 
the disk-outflow system (matter accreting onto compact object with 
considerable intrinsic angular momentum) and summarize the results 
obtained in this case.
In \S 3, we 
present how we model the inflow-outflow system for zero angular 
quasi-spherical accretion. Finally, in \S 4, we discuss some of the 
possible extensions of our work.
\section {\bf Disk-Outflow System:}
\subsection{Formation of CENBOL and the Outflow Geometry}
Before we proceed further, let us describe basic properties of
the rotating inflow and outflow. A rotating inflow with a 
specific angular momentum (angular momentum per unit mass)
entering into a black hole  will have almost constant angular momentum 
close to the black hole for any moderate viscous stress. This is because the
viscous time scale to transport angular momentum 
is generally much longer compared to the infall time scale 
(because near the black hole, the flow `advects' inward with 
enormously large radial velocity) and
even though at the outer edge of the accretion disk 
the angular momentum distribution
may be Keplerian or even super-Keplerian, matter would be highly 
sub-Keplerian close to the
black hole. This happens because the flow {\it has to enter} 
through the horizon
with velocity of light and presence of this large inertial (ram) force,
in addition to usual gravitational and centrifugal forces, makes the
flow sub-Keplerian. This almost constant angular momentum
produces a very strong centrifugal force  which
increases much faster compared to the gravitational force
(because while the centrifugal force varies inversely with the 
{\it qubic} power of the radial distance measured from the 
central gravitating body, gravitational attraction falls
off obeying modified inverse square rule 
{\it because we are using pseudo-Newtonian geometry [41] instead of 
fully general relativistic treatment}) and becomes comparable at 
some specific radial distance, location of which is easy to compute.
Here, (actually, a little farther out, due to
thermal pressure) matter starts
piling up and produces the centrifugal pressure supported boundary layer
(CENBOL). Further close to the black hole, the gravity always wins
and matter enters the horizon supersonically after passing
through a sonic point. CENBOL may or may not have a sharp boundary,
depending on whether standing shocks form or not. Generally speaking,
in a polytropic flow, if the polytropic index $\gamma > 1.5$, then
shocks do not form and if $\gamma <1.5$, only a region of the parameter
space forms the shock [42]. In any case, the CENBOL forms. In this region
the flow becomes hotter and denser 
(matter is either `shock-compressed' or compressed by
the maximization of polytropic pressure of the inflow)
and for all practical purposes
behaves as the stellar atmosphere so far as the formation of
outflows are concerned. 
\footnote{Inflows on neutron stars behave similarly,
except that the `hard-surface' inner boundary condition dictates that the
flow remains subsonic between the CENBOL and the surface rather than becoming
supersonic as in the case of a black hole [43].} 
A part of the hot and dense accreted matter with
shock generate dhigher entropy density (piled up on the CENBOL) 
is then `squirt' as outflow. 
In case where the shock does 
not form,
regions around pressure maximum achieved just outside the inner sonic
point of the {\it inflow} would also drive the flow outwards. In the back of our
mind, we have kind of picture of the outflow 
namely that the outflow is thermally and 
centrifugally accelerated but confined by external pressure of the 
ambient medium.\\
Outflow rates from accretion disks around  
black holes and neutron
stars must be related to the properties of CENBOL which in turn,
depend on the inflow parameters. Subsonic outflows originating 
from CENBOL would pass through sonic points and reach far distances 
as in wind solution.  Figure. 2 represents a schematic diagram showing
the geometry of the disk-jet system proposed in our model. 
The arrows show the axis of the
whirling jet, D(K) stands for the Keplarian part of the disk
and D(SK) stands for the subkeplarian part. CENBOL forms
somewhere inside the D(SK) and J stands for the 
hollow conical jet structure.\\
\begin{center}
\leavevmode
\epsfbox{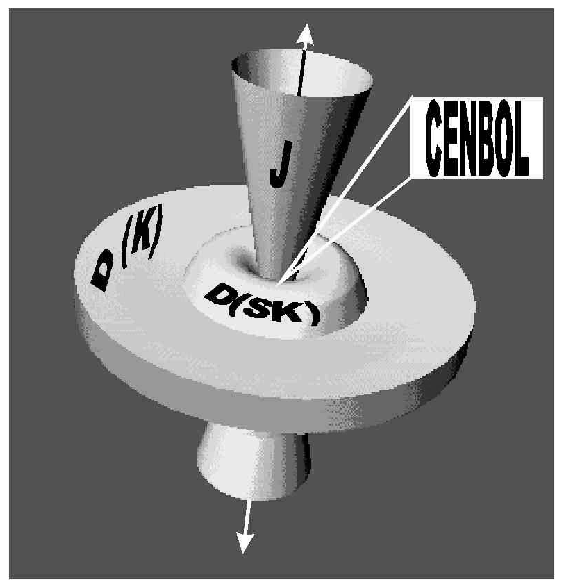}
\end{center}
\begin{center}
{{\bf Fig. 2:}{\small Geometry of the disk-jet system}}\\
\end{center}
There are two surfaces of utmost importance in flows with angular 
momentum. One is the `funnel wall' where the effective potential 
(sum of gravitational potential and the specific rotational energy) 
vanishes. In the case of a purely rotating flow, this is the `zero pressure' 
surface. Flows {\it cannot} enter inside the funnel wall because the pressure 
would be negative. (Fig. 3) The other surface
is called the `centrifugal barrier'. This is the surface
where the radial pressure gradient of a purely rotating flow vanishes and 
is located {\it outside} the funnel wall simply because the flow pressure 
is higher than zero on this surface. Flow with inertial pressure 
easily crosses this `barrier' and either enters
into a black hole or flows out as winds depending on its initial 
parameters (detail classification of the parameter space is in [43]).
In our model the outflow generally hugs the `funnel wall' and goes out 
in between these two surfaces (see [33,36] for detail).\\
\begin{figure}
\vbox{
\vskip -5.0cm
\centerline{
\psfig{file=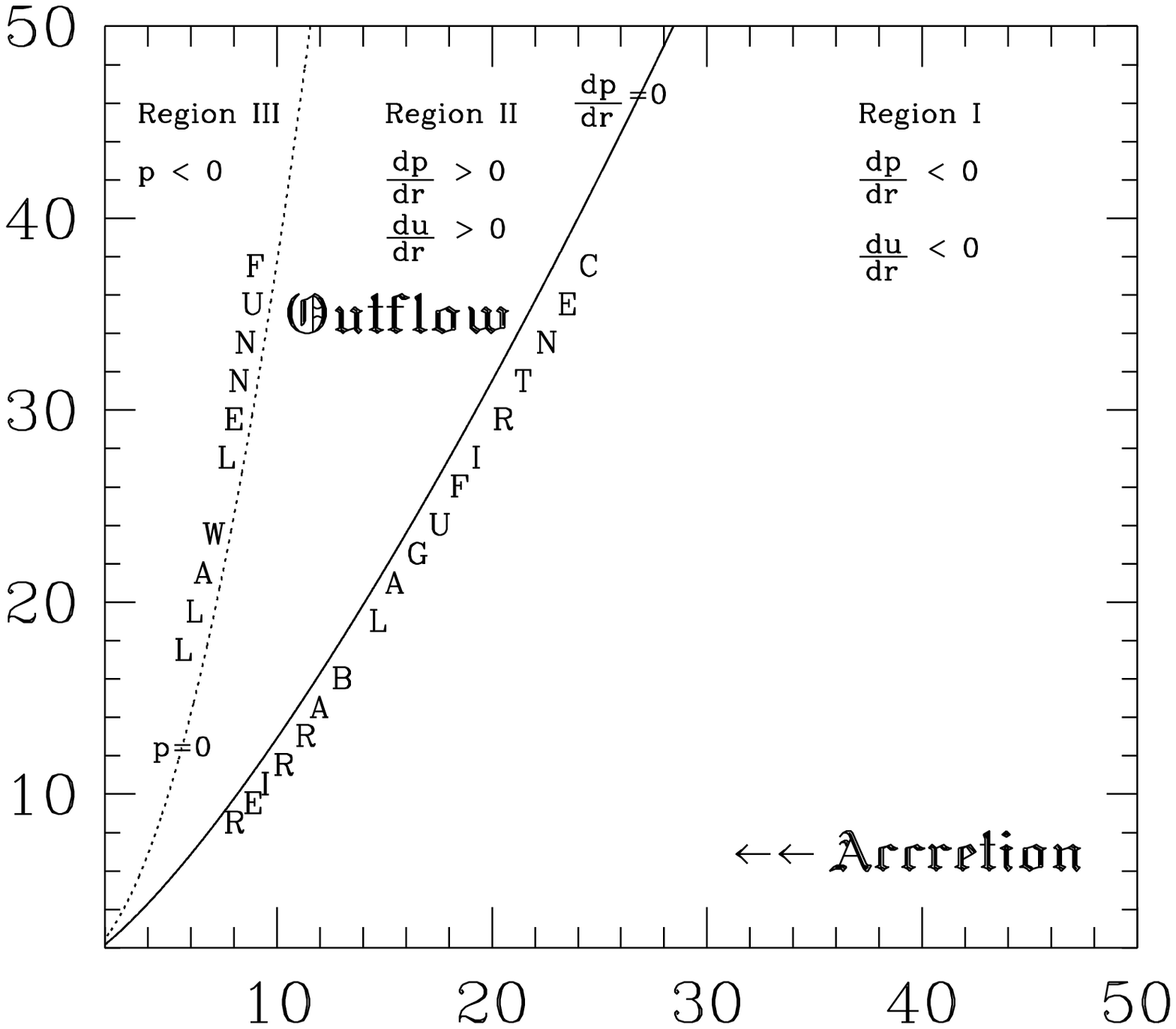,width=10cm}}}
\noindent {{\bf Fig. 3:}
{\small Wind formation from an accretion flow is schematically shown.
The especial properties of centrifugal barrier and funnel wall make
them ideal candidates to collimate
outflows from regions close to the black hole. (Adopted from [36])}}
\end{figure}
\subsection{Model Description, Solution Procedure and Results}
\subsubsection {Model Description}
We consider thin, axisymmetric polytropic inflows in vertical equilibrium
(otherwise known as 1.5 dimensional flow [43]). We
ignore the self-gravity of the flow and viscosity is assumed to be significant
only
at the shock so that entropy is generated. We do the calculations using Paczy\'
nski-Wiita [41]
potential which mimics surroundings of the Schwarzschild black hole.
Considering the inflow to be polytropic, we explore
both the polytropic and the isothermal outflow.\\
For polytropic outflows, the specific energy ${\cal E}$ is assumed to
remain fixed throughout the flow trajectory as it moves from the disk 
to the jet. At the shock, entropy is generated
and hence the outflow is of higher entropy for the same specific energy.\\
For isothermal outflow, we assume that 
the outflow has exactly the {\it same} temperature as that of the
post-shock flow, but the energy is not conserved as matter goes 
from disk to the jet. In other words the outflow is kept in a 
thermal bath of temperature as that of the post-shock flow.
The temperature of the outflow is obtained from the proton 
temperature of the advective region of the disk. The proton temperature
is obtained using the Comptonization, bremsstrahlung, 
inverse bremsstrahlung and Coulomb processes ([44] and references therein).
In both the models of the outflow, we assume that the flow is 
primarily radial. \\
\subsubsection {Solution Procedure}
Let us suppose that matter first enters through the outer sonic point and 
passes through a shock (see [43] for 
parameter space classification). At the shock, part of the incoming matter, 
having higher entropy density is likely to return back as winds through
a sonic point, other than the one it just entered. Thus a combination of 
topologies, one from the region
of accretion and the other from the wind region is required to obtain 
a full solution. In the absence of the shocks,
the flow is likely to bounce back at the pressure maximum of the 
inflow and since the outflow would be heated by photons,
and thus have a smaller polytropic constant, the flow would leave the system
through an outer sonic point different from that of the incoming 
solution. Thus finding a complete self-consistent solution boils 
down to finding the outer sonic point of the outflow and the mass 
flux through it [33-36].\\
For polytropic outflows, the specific energy ${\cal E}$ is assumed to
remain fixed throughout the flow trajectory as it moves from the disk 
to the jet. At the shock, entropy is generated
and hence the outflow is of higher entropy for the same specific energy.
A supply of parameters ${\cal E}$(specific energy of the inflow), 
$\lambda$ (specific angular momentum of the inflow), 
$\gamma$ (polytropic index of the inflow) and 
$\gamma_{o}$ (polytropic index of the {\it outflow}) makes
a self-consistent computation of $R_{\dot m}$ possible (see [33,36]
for detail).
All the physical quantities are measured in the Geometric Unit. 
It is to be noted that when the outflows are produced, one cannot 
use the usual Rankine-Hugoniot relations at the shock location, 
since mass flux is no longer conserved {\it in accretion}, but part of
it is lost in the outflow. Accordingly, we modified the standard 
Rankine-Hugoniot condition [36].\\
\subsubsection{Results}
By simultaneously solving the proper set of equations in appropriate
geometry (see [33,36] for detail), we get the {\it combined} flow 
topologies which is presented as Figure. 4. 
It shows a typical solution which combines the accretion and the outflow.
Mach number (the ratio of the mechanical to the thermal velocity of 
matter) is plotted along the ordinate while the distance measured from
the central object (scaled in the unit of Schwarzschild radius) is plotted
in logarithmic scale along absisca. 
The input parameters are ${\cal E}=0.0005$, ${\lambda=1.75}$ 
and $\gamma=4/3$ corresponding to relativistic inflow. The solid 
curve with an incoming arrow represents the pre-shock region of the inflow 
and the long-dashed curve 
with an arrow inward represents the post-shock
inflow which enters the black hole after passing through the 
inner sonic point (I).
The solid vertical line at $X_{s3}$ (the leftmost vertical transition) with
double arrow represents the shock transition obtained
with exact Rankine-Hugoniot condition (i.e., with no mass loss).
The actual shock location obtained with modified Rankine-Hugoniot condition
[36] is farther out from the original location $X_{s3}$.
Three vertical lines connected with the corresponding dotted curves represent
three outflow solutions for the parameters $\gamma_{o}=1.3$
(top), $1.15$ (middle) and $1.05$ (bottom). The outflow
branches shown pass through the corresponding sonic points. It is
\begin{figure}
\vbox{
\vskip -5.2cm
\hskip 0.0cm
\centerline{
\psfig{figure=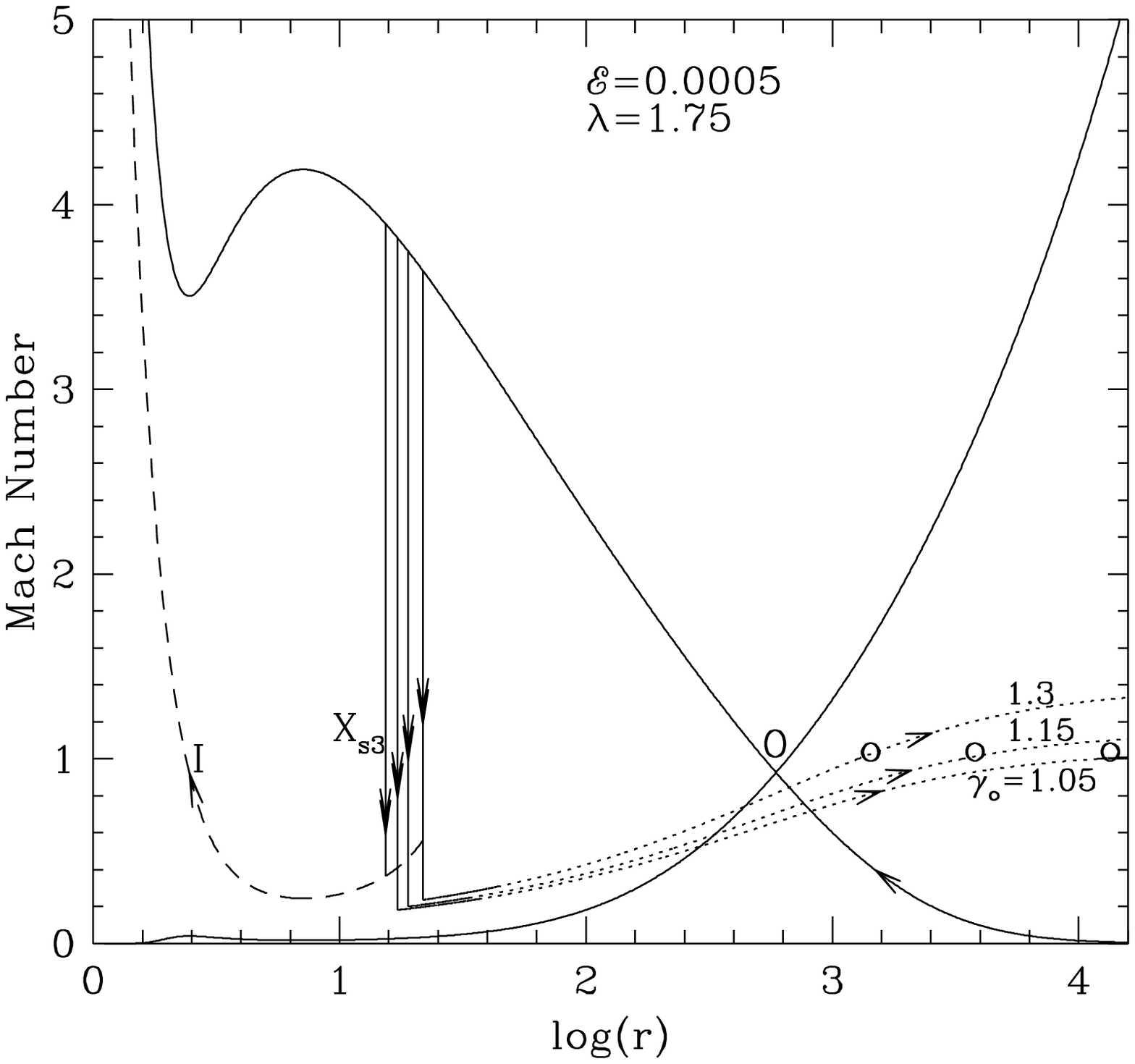,width=10cm}}}
\vskip 0.0cm
\noindent{\small {\bf Fig. 4}:
{\small
Few typical solutions which combine accretion and outflow.
(Adopted from [36])}}
\end{figure}
evident from the figure that the outflow  moves along solution 
curves completely different from that
of the `wind solution' of the inflow which passes through 
the outer sonic point `O'. 
The leftmost shock transition ($X_{s3}$) is obtained
from unmodified Rankine-Hugoniot condition, while the other 
transitions are obtained when the mass-outflow is taken into
account. The mass loss ratio $R_{\dot m}$ 
in these cases are $0.256$, $0.159$ and $0.085$ respectively.\\
We can summarize the results (see [36] for 
details) of our calculation as follows:\\
\noindent a) It is possible that most of the outflows are coming from 
the centrifugally
supported boundary layer (CENBOL) of the accretion disks.\\
\noindent b) The outflow rate generally increases with the proton temperature of
 CENBOL. In other
words, winds are, at least partially, thermally driven. This is reflected more 
strongly
when the outflow is isothermal.\\
\noindent c) Even though specific angular momentum of the flow 
increases the size of the CENBOL,
and one would have expected a higher mass flux in the wind, 
we find that the rate of the
outflow is actually anti-correlated with the $\lambda$ of the inflow. 
On the other hand, if
the angular momentum of the outflow is reduced, we find that the rate of
 the outflow is correlated with
$\lambda$ of the outflow. This suggests that the outflow is 
partially centrifugally driven as well.\\
\noindent d) The ratio $R_{\dot m}$ is generally anti-correlated with the inflow
 accretion rate. That is,
disks of lower luminosity would produce higher $R_{\dot m}$.\\
\noindent e) Generally speaking, supersonic region of the inflow do not have pre
ssure maxima. Thus,
outflows emerge from the subsonic region of the inflow, whether the shock actual
ly forms or not.\\
If we introduce an extra radiation pressure term (with a term like
$\Gamma/r^2$ in the radial force equation, where $\Gamma$ is the contribution
due to radiative process), particularly important for
neutron stars, the outcome is significant. In the 
{\it inflow}, outward radiation
pressure weakens gravity and thus the shock is located farther out. 
The temperature is cooler and therefore the outflow rate is lower.
If the term 
is introduced only in the
{\it outflow}, the effect is not significant [36]. However, we understand
that inclusion of only $\frac{\Gamma}{r^2}$ term does not give the
whole picture of the various radiative processes taking place in the
disk-jet system and  a more general and exact
form of the radiative force term is to be included in the set of
equations governing the inflow-outflow system.\\
An interesting situation arises when the polytropic index of the 
outflow is large
and the compression ratio of the flow is also very high. 
In this case, the flow virtually
bounces back as the winds and the outflow rate can be equal to the
inflow rate or even higher, thereby evacuating the disk. 
In this range of parameters, most, if not all,
of our assumptions may breakdown completely because the situation 
could become inherently time-dependent.
It is possible that some of the black hole systems, including that 
in our own galactic center, may have undergone such evacuation phase 
in the past and gone into quiescent phase (see [33-36] for detail).\\
Strong winds are suspected to be present in Sgr $A^*$ at our galactic
center [45-46]. We have shown that when the inflow rate itself
is low (as in the case for Sgr $A^*$; $\sim 10^{-3} - 10^{-4}$
Eddington rate), the mass outflow rate is very high, almost to the point
of evacuating the disk. This prompted us to strongly speculate that
the spectral properties of our galactic center could be explained
by inclusion of winds using our model [36].\\
\section{Outflow from Bondi Type Accretion}
\subsection{In Search for a Suitable Surface}
For some black hole models of active galactic 
nuclei, inflow may not have accretion disk [47].
Accretion is then quasi-spherical having almost zero or negligible angular 
momentum (Bondi [48] type accretion).
Absence of angular momentum rules out the possibility of 
formation of the 
Rankine-Hugoniot shock as well as the 
polytropic pressure maxima. So, for quasi-spherical accretion,
CENBOL formation (as discussed in the earlier section)
is not possible. 
It has been shown that [49-50]
for quasi-spherical 
accretion onto 
black holes, steady state situation may be developed 
close to the black hole where a standing 
collisionless shock may form due to the plasma instabilities and for
nonlinearity introduced by small density perturbation. 
This is because, 
after crossing the sonic point
 the infalling matter (in plasma form) becomes highly supersonic.
Any small perturbation and slowing down of the infall velocity
will create a piston and produce a shock. A spherically symmetric shock 
produced in such a way will accelerate a fraction of the inflowing
plasma to relativistic energies. The shock accelerated 
relativistic particles 
suffer essentially no Compton loss and are assumed to lose energy
only through proton - proton ($p - p$) collision. 
These relativistic hadrons are not readily captured by the 
black hole [51] rather
considerable high energy density of these
relativistic protons would be maintained to support a standing, collisionless,
spherical shock around the black hole [50].
Thus, a self-supported standing shock may be produced {\it even} for accretion
with zero angular momentum.
In this work, we take this 
pair-plasma pressure mediated shock surface as the alternative
of the CENBOL which can be treated as the  effective physical hard
surface which, in principle  mimics the ordinary stellar surface regarding 
the mass outflow.\\
The condition necessary for the development and maintenance of such a 
self-supported spherical shock is 
satisfied for the high Mach number solutions [52].
Keeping this in the back of our mind, for our present work,
we concentrate only on low energy accretion to obtain high shock Mach number.
Considering low energy 
(${{\cal E} \lsim  0.001}$) accretion, 
we assume that particles accreting
toward black hole are shock accelerated via first order Fermi
acceleration 
producing relativistic protons. Those relativistic protons
usually scatters several times before being captured by the black hole.
These energized particles, in turn, provide sufficiently outward
pressure to support a standing,
collisionless shock. A fraction of the energy flux of infalling matter
is assumed to be converted into radiation at the shock standoff
distance through hadronic ($p -p$) collision and mesonic ($\pi^{\pm},\pi^0$) 
decay. Pions generated by this process, decay into relativistic electrons,
neutrinos/antineutrinos and
produces high energy $\gamma$ rays (see [38] for details).
These electrons produce the observed non-thermal radiation by Synchrotron
and inverse Compton scattering. The
overall efficiency of this mechanism depends largely on the shock location.
Luminosity produced by this fraction is used to obtain the shock 
location for the present work. \\
\subsection {At the end of the Search}
At the shock surface,
density of the post-shock material shoots up and velocity falls
down, infalling matter starts piling up on the shock surface. The post
shock relativistic hadronic pressure then gives a kick to the piled up
matter the result of which is the ejection of outflow from the shock surface.
For this type of inflow, accretion is known to proceed smoothly
after a shock transition, since successful subsonic solutions
have been constructed for accretion onto black holes
embedded within normal stars with the boundary condition
${u = c}$; where $u$ is the infall velocity of matter and $c$ is the
velocity of light in vacuum.
The fraction of energy converted, the shock compression ratio
${R_{comp}}$,
along with the ratio of post shock relativistic
hadronic pressure to infalling ram pressure at a given shock location
are obtained from the steady state shock solution of 
Ellision and Eichler [52-53].
The shock location as a function of the specific energy ${\cal E}$ of the
infalling matter and accretion rate is then self-consistently obtained
using the above mentioned quantities.
We then calculate the amount of mass outflow rate $R_{\dot m}$ from the 
shock surface using combination of exact transonic inflow outflow
solutions and study the dependence of $R_{\dot m}$ on various physical
entities governing the inflow-outflow system [38].\\
\subsection{Model Description, Solution Procedure and Results}
\subsubsection{Model Description and Solution Procedure}
We assume that a Schwarzschild type black hole
quasi-spherically accretes low energy (${\cal E} \lsim 0.001$) fluid
obeying polytropic equation of state. 
We also assume that the accretion rate 
with which the fluid is being accreted, is not a function
of $r$ ($r$ being the radial distance measured from the 
central object scaled in the unit of 
Schwarzschild radius). For simplicity of calculation, we choose geometric unit
to measure all the relevant quantities. We ignore the
self-gravity of the flow and the calculation is being done using
Paczy$n^{\prime}$ski-Wiita [41]
potential which mimics surrounding of the Schwarzschild
black hole. 
As already mentioned, we assume that a steady, collisionless shock forms
at a particular distance (measured in the unit of Schwarzschild radius) due
to the instabilities in the plasma flow. We also assume that for our model,
the effective thickness of the shock is small enough compared
to the shock standoff distance.
For simplicity of calculation, we assume that the outflow is also
quasi-spherical.
\begin{figure}
\vbox{
\vskip -5.0cm
\centerline{
\psfig{file=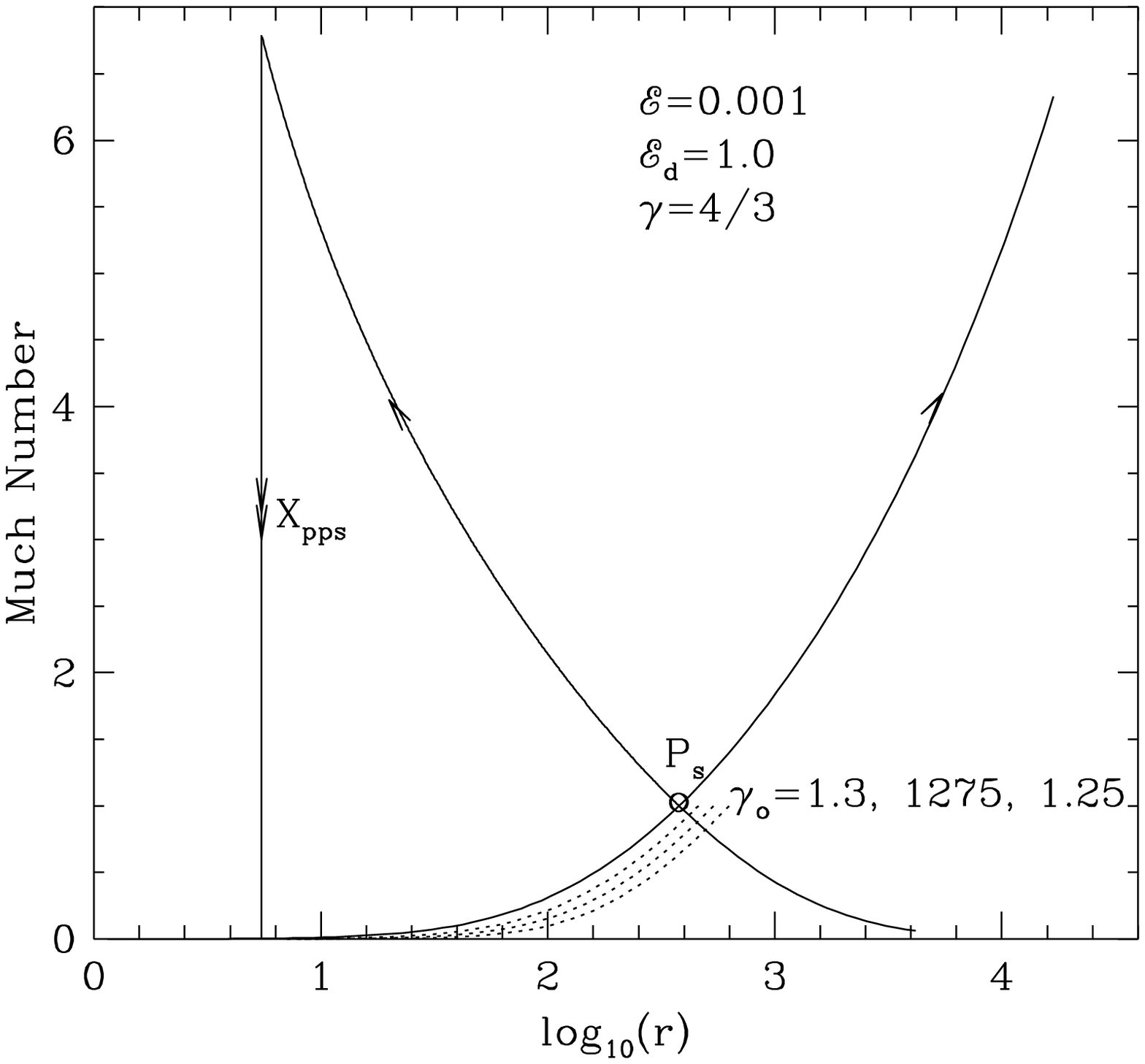,width=10cm}}}
\noindent {{\bf Fig. 5:}
{\small 
Solution topology for three different $\gamma_o$ (1.3, 1.275, 1.25)
for ${\cal E}$ = 0.001, ${\dot M}_{in}$ = 1.0, ${\cal E}_d$ = 1.0, $\gamma = \frac
{4}{3}$.
$P_s$ indicates the sonic point of the {\it inflow} where $X_{pps}$ stands for the
shock location. (Adopted from [38]). See text for details.}}
\end{figure}
It is obvious from the above discussion
that ${R}_{\dot m}$ should have some complicated functional dependences on
the inflowing parameters through the shock location (see [38] for detail).\\
\subsubsection{Results}
By simultaneously solving the proper set of equations in appropriate
geometry (see [38] for detail), we get the {\it combined} flow
topologies which is presented as Figure. 5.
It shows a typical solution which combines the accretion and the outflow.
Mach number (the ratio of the mechanical to the thermal velocity of
matter) is plotted along the ordinate while the distance measured from
the central object (scaled in the unit of Schwarzschild radius) is plotted
in logarithmic scale along absisca.
The input parameters are ${\cal E}$ = 0.001, ${\dot M}_{in}$
=1.0 Eddington rate (${\cal E}_d$ stands for the Eddington rate
in the figure)  and $\gamma = \frac{4}{3}$
corresponding to relativistic inflow.
The solid curve with an arrow represents the pre-shock region of the
inflow and the solid vertical line with double arrow at $X_{pps}$
(the subscript $pps$ stands for $p$air $p$lasma mediated
$s$hock) represents
the shock transition. Location of shock is obtained using the eqs.(4)
for a particular set if inflow parameters mentioned above. Three
dotted curves show the three
different outflow branches corresponding to different polytropic
index of the outflow as $\gamma_o$ = 1.3(left most curve),
1.275 (middle curve) and 1.25(rightmost curve).
It is evident from the figure that the outflow moves along the
solution curves completely different from that of the
``wind solution" (solid line marked with an outword directed arrow)
of the inflow which passes through the sonic point $P_s$. The mass loss
ratio $R_{\dot m}$ for these cases are 0.0023, 0.00065 and 0.00014
respectively.\\
We can summarize our results (see [38] for details) 
obtained in this case as follows:\\
\noindent a)It is possible that outflows for quasi-spherical
Bondi type accretion onto a Schwarzschild black hole are coming from the
pair plasma pressure mediated shock surface.\\
\noindent b) The outflow rate monotonically increases with the 
specific energy of the inflow and nonlinearly increases with the 
Eddington rate of the infalling matter.\\
\noindent c) $R_{\dot m}$, in general, correlates with $\gamma_o$ 
but anticorrelates with $\gamma$.\\
\noindent d) Generally speaking, as our model deals with high shock 
Mach number (low energy accretion) solutions, outflows in our work 
always generate from the supersonic branch of the inflow, i,e, 
shock is always located {\it inside} the sonic point.\\
\noindent e) Unlike the mass outflow from the disk-outflow case
around black holes [33-36]
here we found [38] that the value 
of $R_{\dot m}$ is distinguishably small. This is
because matter is ejected out due to the pressure of the relativistic
plasma pairs which is {\it very much less} in comparison to the pressure
generated due to the presence of significant angular momentum.
However, in the present work we have dealt only high Mach number
solution which means matter is accreting with very low energy (`cold inflow',
as it is described in literature). This is another possible reason to
obtain a low mass loss rate. If, instead of high Mach number solution,
we would use low Mach number solution, e,g, high energy accretion, the
mass outflow would be considerably higher (this is obvious because
it has already been established in present work that $R_{\dot m}$
increases with ${\cal E}$ (see [38] for detail).
\section{\bf Future Perspectives}
Based on the works have already been done, our goals for the future work
are essentially the following:\\
\begin{enumerate}
\item We have carried out our calculations for the Schwarzschild black
hole using Paczy$n^{\prime}$ski-Wiita [41] pseudo-Newtonian potential. 
Now we would like
to extend our calculations in fully general relativistic framework so
that the parameter space for calculation gets modified and the mass-loss
rate calculated in this way could be compared with our previous results.
\item In our work, we assumed that the magnetic field is absent.
Magnetized winds from the accretion disks have so far been considered
in the context of a Keplarian disk and not in the context of
sub-Keplarian flows on which we concentrate here. It is not unreasonable
to assume that our prime surface for the wind formation, the CENBOL
would still form when magnetic fields are present and since the Alfven
speed is, by definition, higher compared to the sound speed, the
acceleration would also be higher than what we computed here. Morever,
introduction of toroidal magnetic field  in our model 
with its associated ``hoop"
stress, would lead to a better understanding of the ``collimation
problem" of the jets.
\item Recently it has been suggested ([54] \& references therein) that
significant nucleosynthesis is possible in the centrifugal pressure
supported dense and hot region of the accretion flow which deviate from
the disk around the black holes. Attempts had been made to compute the
composition changes and energy generation due to such nuclear processes
as a function of the radial distance from the black hole. We suggest that
the outflows produced from this region would carry away modified
composition  and contaminate the atmosphere of the surrounding stars
and the galaxies in general. Unlike the present calculation, where the
outflow consists of $m_p$ only (proton jet), now we will be trying to
take the weighted average of the heavier elements produced by the
nucleosynthesis in advective accretion disks as the constituent
elements of the outflow. 
\item Finally, we would like to carry out all of our calculations (done
for Schwarzschild black hole) in Kerr space-time to bring the whole
picture into focus.\\
\end{enumerate}
{\large\bf Epilogue}\\ \\
\noindent
It is to be noted that although the existence of astrophysical outflows
and jets from the galactic and extragalactic sources 
are well known, their rates are not. Similarly, till date,
there is no definitive model present in the literature which can handle
the origin of this outflow in a self-consistent way. Hence we think that
the formation and dynamics (acceleration and collimation) 
of these outflows are the 
open problems in present day theoretical astrophysics. Along with our
present analysis of mass outflow in Schwarzschild geometry, if we can
carry out our calculation in Kerr geometry as well, we 
strongly believe that these combined
calculations definitely could shed some new light on the origin and
energetics of the astrophysical jets.\\ \\
\noindent
{\large\bf Acknowledgements}\\ \\
\noindent
The author expresses his gratitude to Prof. S. K. Chakrabarti
for introducing him to this subject and for his 
guidance, to Dr. S. 
Chakrabarti, and Mr. A. Ray
of his Institute for carefully going through the manuscript
and for providing valuable suggestion to improve the quality of this
article.
He is also thankful to Prof. A. R. Rao of 
Tata Institute of Fundamental research for constructive interaction 
during the YATI conference. The present article is basically an
organized summary of a couple of seminars and colloquiums
presented by the author at various Institutes and conferences
during the period November 1998 to March 1999. 
The author was greatly benefited from 
discussions with several faculty members, post doctoral fellows and 
research students of the Institutes he visited and with some of 
the participants of last Texas Symposium (Dec. 1998) 
held in Paris during his 
participation. Among them,
special thanks are for Prof. P. A. Charles, Dr. T. L. Grey and
Dr. P. Saha from Dept. of Astrophysics, University of Oxford,
Prof. W. Kundt from Institute f$\ddot{u}$r
Astrophysics, University of Bonn, Prof. P. 
Biermann, Dr. H. Falke, Dr. T. Ensllin 
and Dr. Yiping Yang of Max Plack Institute for Radio Astronomy,
Prof. B. Czerney and Ms. A. Rozanska from Nicolaus Copernicus Astronomical
Centre, Warsaw, Poland, Prof. M. Begelman from ZILA, Colorado, Prof.
M. Ostrowski from Obserwatorium Astronomiczne, Krakow, Poland,
Prof. J. V. Narlikar, Prof. A. Kembhabi, Dr. S. Bose, Dr. R. Misra
and Dr. R. Srianad from Inter University Centre for Astronomy and
Astrophysics and Prof. Gopal Kriskna from National Centre for
Radio Astronomy. 


\section*{References}
\begin{thebibliography}{}
\bibitem{ } De Young, S. D. (1991) {\it Science} {\bf vol.252} p. 389
\bibitem{ } Heyvaerts, J. (1999) {\it Proc. 19th Texas Symposium
in Relativistic Astrophysics \& Cosmology} ({\bf In Press})
\bibitem{ } Begelman, M. C. et. al. (1984) {\it Rev. Mod. Phys.} 
{\bf Vol. 56, No. 2, Part 1} p. 255
\bibitem{ } Mirabel, I. F. \& Rodriguez, L. F. (1998) {\it Nature} {\bf Vol. 392}
p. 673
\bibitem{ } Jennison, R. C. \& Das Gupta, M. K. (1953) {\it Nature}
{\bf 172} p. 996
\bibitem{ } Bridle, A. H. \& Perley, R. A. (1984) {\it ARA \& A} {\bf 22} 
p. 319
\bibitem{ } Blandford, R. D. (1990) in {\it Active Galactic Nulei,
Saas-Fee Advance Course 20} p. 161
\bibitem{ } Hughes, P. A. (ed.) (1991) {\it Beams and Jets in
Astrophysics} {\bf CUP} 
\bibitem{ } Baade, W. \& Minkowski, R. (1954) {\it Ap. J.} {\bf 119} p.215
\bibitem{ } Margon, B. (1984) {\it ARA\&A} {\bf 22} p. 507
\bibitem{ } Vermeulen, R. (1993) in {\it Astrophysical Jets}, 
Burgarella, D. et. al. (ed). p. 241
\bibitem{ } Rees, M. J. (1971) {\it Nature} {\bf 229} p. 312
\bibitem{ } Rees, M. J. (1984) {\it ARA\&A} {\bf 22} p. 471
\bibitem{ } Kafatos, M. (ed) (1988) {\it Supermassive Black Holes}
{\bf CUP}
\bibitem{ } Ferrari, A. (1998) {\it ARA \& A} {\bf 36} p. 539
\bibitem{ } Madejski, G. M. (1999) in {\it Theory of Black Hole 
Accretion Disks}, Abramowicz, G. et. al. (ed.) ({\bf In Press})
\bibitem{ } Falke, H. (1994) {\it Starved Holes and Active Neuclei: 
The Central Engine in Galactic Centers}, {\bf Ph.D. Thesis}. 
\bibitem{ } Ferrari, A. Trussoni, E., Rosner, R. \& Tsinganos, K. (1985) 
{\it Ap. J} {\bf 294} p. 397
\bibitem{ } Fukue, J. (1982) {\it PASJ} {\bf 34} p. 163
\bibitem{ } Contopoulos, J. (1995), {\it Ap. J} {\bf 446} p. 67
\bibitem{ } Chakrabarti, S. K. (1986) {\it Ap.J} {\bf 303} p. 582
\bibitem{ } Pelletier, G. \& Pudritz, R. E. (1992) {\it Ap. J.} {\bf 394}
p. 117
\bibitem{ } K$\ddot{o}$nigl, A. (1989) {\it Ap. J}, {\bf 342}, p. 208 
\bibitem{ } Chakrabarti, S. K. \& Bhaskaran, P. (1992) {\it Mon. Not. Roy. 
Soc.} {\bf 255} p. 255
\bibitem{ } Hawley, J.W., Smarr, L. \& Wilson, J. (1984) {\it Astrophys. J.}
{\bf 297} p. 296.
\bibitem{ } Hawley, J.F., Smarr, L.L., \&  Wilson, J.R. (1985)
{\it Astrophys. J.} {\bf 55} p. 211
\bibitem{ } Eggum, G. E., Coroniti, F. V., Katz, J. I. (1985) 
{\it Astrophys. J.} {\bf 298, L} p. 41
\bibitem{ } Molteni, D., Lanzafame, G. \& Chakrabarti, S. K. (1994)
{\it  Astrophys. J.} {\bf 425} p. 161
\bibitem{ } Molteni, D., Ryu, D. \& Chakrabarti, S.K. 
(1996) {\it Astrophys. J.} {\bf 470} p. 460
\bibitem{ } Ryu, D., Chakrabarti, S. K., \& Molteni, D. (1997)
{\it  Astrophys. J.} {\bf 474} p. 378
\bibitem{ } Nobuta, K. \& Hanawa, T. (1999) {\it Ap. J.} {\bf 510} p. 614
\bibitem{ } Chakrabarti, S. K. (1998) in {\it Observational Evidence
for Black Holes in the Universe, Ed. S.K. Chakrabarti 
(Kluwer Academic: Holland)} p. 19
\bibitem{ } Das, T.K. (1998) in {\it Observational Evidence
for Black Holes in the Universe, Ed. S.K. Chakrabarti 
(Kluwer Academic: Holland)} p. 113
\bibitem{ } Das, T.K. (1998) in {\it Proc-. X-th International 
Summer School Seminar on Recent Problems in Theoretical and 
Mathematical Physics, Kazan} ({\bf In Press})
\bibitem{ } Das, T.K. \& Chakrabarti, S. K. (1999) in 
{\it Proc. 19th Texas Symposium in 
Relativistic Astrophysics \& Cosmology (CD ROM Version).} 
\bibitem{ } Das, T. K. \& Chakrabarti, S. K. (1999) Submitted
to {\it Cls. Qunt. Grav.} 
\bibitem{ } Das, T.K. (1999) {\it Ind. Jour. Phys.} {\bf 73B(1)(Rapid
Communication)} Pg. 1-7.
\bibitem{ } Das, T.K. (1999) {\it Monthly Notices of the Royal
Astronomical Society} {\bf In Press}
\bibitem{ } Das, T.K. (1999) in {\it Proc. 19th Texas Symposium on
Relativistic Astrophysics \& Cosmology (CD ROM Version).} 
\bibitem{ } Das, T.K. (1998) in {\it Proc.- International Mini 
Workshop on Applied Mathematics.} ({\bf In Press})
\bibitem{ } Paczyn\'ski, B. \& Wiita, P. J. (1980) {\it A \& A} {\bf 88}
p. 23
\bibitem{ } Chakrabarti, S. K. (1996)  {\it Ap. J} {\bf 464} p. 664
\bibitem{ } Chakrabarti, S. K. (1989) {\it Ap. J} {\bf  347} p. 365
\bibitem{ } Chakrabarti, S. K. (1997) {\it  Ap. J} {\bf 484} p.  313
\bibitem{ } Genzel, R. et. al. (1994) {\it Rep. Prog. Phys.}
{\bf  57} p. 417
\bibitem{ } Eckart, A. \& Genzel, R. (1997) {\it MNRAS.} {\bf  284} 
p. 576
\bibitem{ } Rees, M. J. (1977) {\it Quart. Jour. Roy. Astr. Soc.}
{\bf 18} p. 429
\bibitem{ } Bondi H. (1952) {\it MNRAS} {\bf 112} p. 195
\bibitem{ } Meszaros P. \& Ostriker J P. (1983) 
{\it Ap. J} {\bf 273L} p. L59
\bibitem{ } Kazanas D. \& Ellisopn D C. (1986) {\it Ap. J} {\bf 304} p. 178
\bibitem{ } Protheroe R J. \& Kazanas D. (1983) {\it Ap. J.} {\bf 265}  
p. 620
\bibitem{ } Ellision D C. \& Eichler D. (1984) {\it Ap. J.} {\bf 286} p. 691
\bibitem{ } Ellision D C. \& Eichler D. (1985) {\it Phys. Rev. D}
{\bf 55} p. 2735
\bibitem{ } Chakrabarti, S. K. \& Mukhopadhyay, B. (1999) {\it A \& A}
({\bf In Press})
\end{thebibliography}
\end{document}